\documentclass[aps, pre, amsmath, amssymb, amsfonts, twocolumn, showpacs, floatfix]{revtex4}
\usepackage{graphicx}
\usepackage{subfigure}
\usepackage{dcolumn}
\usepackage{bm}

\newcommand     {\beq}[1]         { \begin{equation} #1 \end{equation} }
\newcommand     {\beqa}[1]        { \begin{eqnarray} #1 \end{eqnarray} }

\newcommand     {\eone}             { \epsilon_1 }
\newcommand     {\etwo}             { \epsilon_2 }
\newcommand     {\eonemin}          {\epsilon_1^{\textrm{min}}}
\newcommand     {\eonemax}          {\epsilon_1^{\textrm{max}}}
\newcommand     {\etwomin}          {\epsilon_2^{\textrm{min}}}
\newcommand     {\etwomax}          {\epsilon_2^{\textrm{max}}}

\begin{document}

\title{A simple beam model for the shear failure of interfaces}

\author{ F.\ Raischel${}^1\footnote{Electronic 
address:raischel@ica1.uni-stuttgart.de}$, F.\ Kun${}^{1,2}$, and H.\
J.\ Herrmann${}^{1}$} 
\affiliation{
${}^1$ ICP, University of Stuttgart, Pfaffenwaldring 27 , D-70569 Stuttgart, Germany\\
${}^2$Department of Theoretical Physics, University of Debrecen, P.\
O.\ Box:5, H-4010 Debrecen, Hungary }

\date{\today}
             
\begin{abstract}
We propose a novel model for the shear failure of a glued interface
between two solid blocks. We model the interface as an array of
elastic beams which experience stretching and bending under shear load and
break if the two deformation modes exceed randomly distributed
breaking thresholds. The two breaking modes can be independent or
combined in the form of a von Mises type breaking criterion. Assuming global
load sharing following the beam breaking, we obtain analytically the
macroscopic constitutive behavior of the system and describe the
microscopic process of the progressive failure of the interface. We
work out an efficient simulation technique which allows for the study
of large systems. The limiting case of very localized interaction of
surface elements is explored by computer simulations. 

\end{abstract}  
\pacs{02.50.-r,05.90.+m, 81.40.Np}
\maketitle
\section{Introduction}
Solid blocks are often joined together by welding or
glueing of the interfaces which are expected to sustain various types
of external loads. 
When an elastic interface is subjected to an increasing load applied
uniformly in the perpendicular direction, in the early stage of the 
failure process cracks nucleate randomly along the interface.
Due to the heterogeneous microscopic properties of the glue,
these cracks can remain stable under increasing load, which results
in a progressive damage of the interface. This gradual softening process
is followed by the localization of damage which leads then to the global
failure of the interface and separation of the two solid blocks.  

Interfacial failure plays a crucial role in fiber reinforced
composites, which are constructed by embedding fibers in a matrix
material  \cite{cahn}. 
Composites are often used as structural components 
since they have very good mass specific properties, {\it i.e.} they
provide high strength with a relatively low mass, 
preserving this property even under extreme conditions. 
The mechanical performance of composites is mainly determined by the
characteristic quantities of the constituents (fiber and matrix), 
and by the fabrication process which controls the material's
microstructure,  the formation of damage prior to
applications, and the properties of the fiber-matrix interface. In
many cases the reinforcement is  a 
unidirectional arrangement of long fibers resulting in highly anisotropic
mechanical properties, {\it i.e.} in the direction of the fiber axis the
composite exhibits high strength and fracture toughness since the
load is mainly carried by fibers, however, in the perpendicular
direction the load bearing capacity is provided solely by the matrix
material. Hence the dominant failure mechanism  of unidirectional composites
perpendicular to the fibers' direction is  shear. Failure
here occurs mainly due to the debonding of the fiber-matrix interface. 

Since disordered properties of the glue play a crucial role in the
failure of interfaces, most of the theoretical studies in this field
rely on discrete models \cite{hh_smfdm,hh_cdm} which are able to
capture heterogeneities and 
can account for the complicated interaction of nucleated cracks.
The progressive failure of glued interfaces
under a uniform load perpendicular to the interface has recently been
studied by means of  fiber bundle models
\cite{daniels+proc_rsa+1945,sornette_jpa_1989,kloster_pre_1997,raul_varint_2002,raul_burst_contdam,moreno_fbm_avalanche,chakrabarti_fatigue,chakrabarti_phasetrans}. 
Several aspects of the failure process have been revealed such as the 
macroscopic constitutive behavior, the distribution of avalanches of
simultaneously failing glue and the structure of failed glue regions
\cite{batrouni_intfail_pre_2002}. Considering a hierarchical scheme for the load redistribution
following fiber failure, a cascading mechanism was proposed for the
softening interface in
Ref.~\cite{delaplace_ijss_1999,roux_damint_physicaa_1999}. 
The roughness of the crack front propagating between two rigid plates
due to an opening load was studied in the framework of the fuse
model. The microcrack nucleation ahead the main crack and the
structure of the damaged zone were analyzed in detail \cite{zapperi_crackfuse_eujb_2000}. 
The shear failure of an interface between two rigid blocks has very
recently been investigated by discretizing the interface in terms of
springs. It was shown that shear failure of the interface occurs as a
first order phase transition \cite{knudsen_breaksurf_condmat_2004}.

 In the present paper we study the shear failure of the glued
interface connecting  two solid blocks in the framework of a novel type of
model. In our model the interface is discretized in terms of elastic
beams which can be elongated and bent when exposed to shear
load. Breaking of a beam is caused by two breaking modes, {\it i.e.}
stretching and bending, characterized by randomly distributed
threshold values. The two breaking modes can be either independent or
combined in terms of a von Mises type breaking criterion
\cite{hjh_prb_fractdis_1989}. 
Assuming long range interaction among the beams, we obtained the full
analytic solution of the model for the macroscopic
response of the interface, and for the microscopic process of
failure. We show that the presence of two breaking modes lowers the
critical stress and strain of the material without changing the
statistics of bursts of simultaneously failing elements with respect
to the case of a single breaking mode. The coupling of breaking modes
results in further reduction of the strength of the interface. We
demonstrate that varying the relative importance of the two breaking
modes the macroscopic response of the interface can be tuned over a
broad range. The limiting case of very localized interaction of beams
is also considered. We determine the constitutive behavior and the
distribution of avalanches of breaking beams for the case when beams
interact solely with their nearest and next-nearest neighbors in a
square lattice.
An effective simulation
technique is worked out which makes it possible to study systems of large
size. 

\section{Properties of the model}\label{sec:model}

\begin{figure}[h]
    \subfigure[]{
      \includegraphics[clip, scale=0.45]{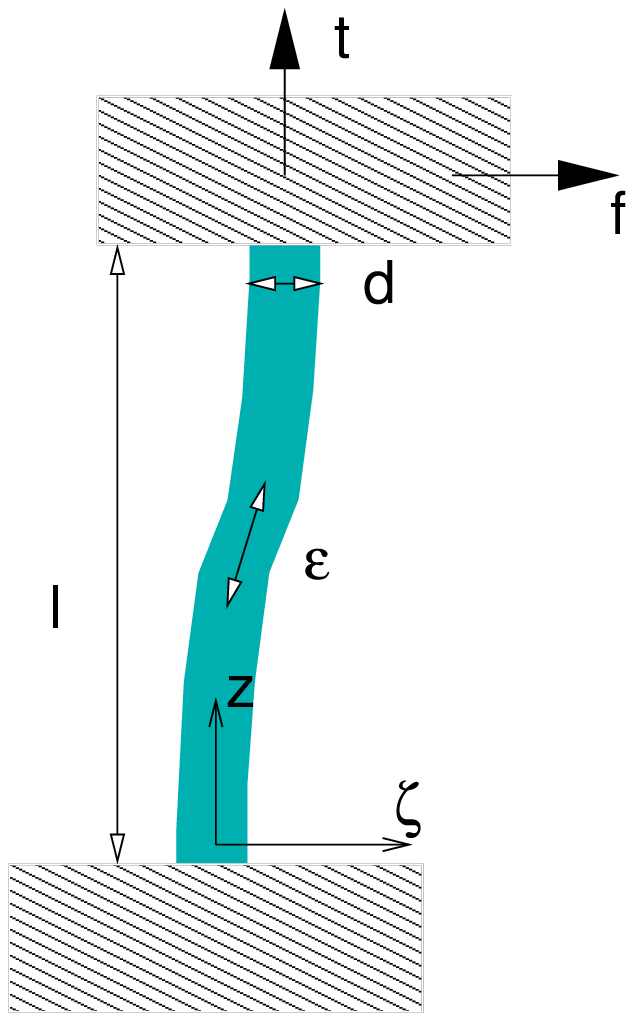}
      \label{gr:theory:bendsketch:a} 
    } \hfill %
    \subfigure[]{
      \includegraphics[clip, scale=0.45]{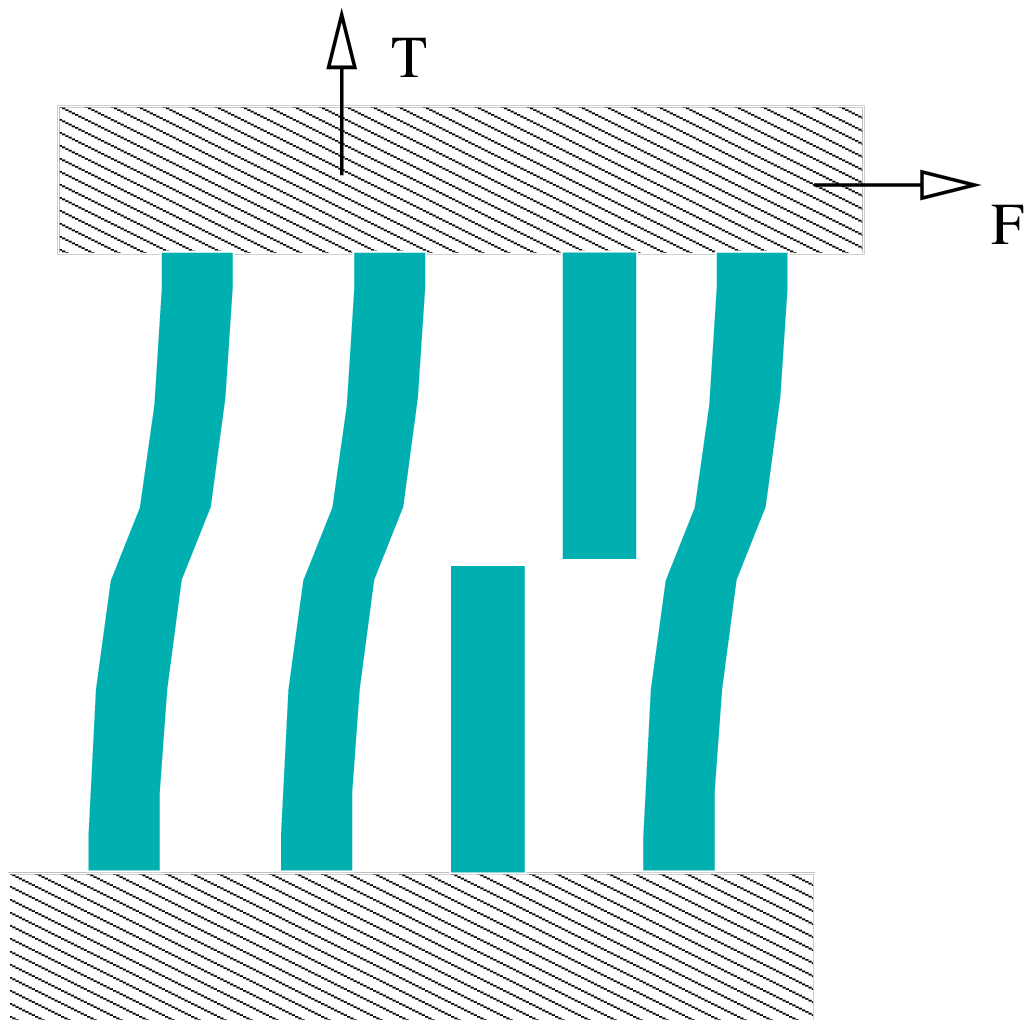}
      \label{gr:theory:bendsketch:b} 
    }
    \caption{(a) Shearing of a single beam between two rigid
      plates. Since the distance $l$ between the plates is kept
      constant, the beam experiences stretching and bending
      deformation, with longitudinal  $t$ and shear $f$ forces. (b)
      Shearing of an array of beams, with the corresponding forces. In
      the case shown, one beam is broken.}
    \label{gr:theory:bendsketch} 
\end{figure}

In our model we represent the glued interface of two solid blocks as
an ensemble of parallel beams connecting the two surfaces.  
First, we derive an analytical description of a single beam of quadratic
cross section clamped at both ends and sheared by an external force
$f$, see Fig.~\ref{gr:theory:bendsketch:a}. The shearing is exerted in
such a 
way that the distance $l$ between the two clamping planes is kept
constant. Consequently, the beam experiences not only a torque
$m$, but also a normal force $t$  due to the elongation $\Delta l$,
which is characterized by the longitudinal strain $\epsilon = \Delta l /
l$.  

We derive the form of the deflection curve of the beam, as well as
the magnitude of the tension force. It is necessary to
introduce some approximations, so the model can be incorporated into
the simulation code in a sensible way.  
Following the procedure outlined e.~g. in \cite{ll_elast_en}, we
solve the differential equation for the beam deflection $\zeta (z)$
under the influence of the lateral force $f$ and a given stretching
force $t$. We then solve self-consistently for $t(f)$, with $t$
being the result of the longitudinal elongation. 

The governing differential equation for the bending situation depicted
in Fig.~\ref{gr:theory:bendsketch:a} can be cast in the form 
\beq{ \label{eq:theory:beam_forces}
  \zeta'''(z) - \frac{t}{E I}  \zeta' (z) = -\frac{f}{E I},
  }
with boundary conditions
\begin{eqnarray}
  &\zeta(0) &= 0, \nonumber \\
  &\zeta' (0) &= 0, \nonumber \\
  &\zeta'' (l/2) &= 0.
\end{eqnarray}
Here, $E$ denotes the modulus of elasticity, and $I$ is the moment of
inertia for bending of the beam. For a beam of rectangular
cross-section, we have $I = d^4 /12$, where $d$ is the side length.  
Let us briefly motivate this ansatz by stating that the second
derivative $\zeta''(z)$ is proportional to the torque on the beam, so
consequently it needs to vanish at the beam half-length
$l/2$. Accordingly, the third derivative $\zeta'''(z)$ is proportional
to the shearing force exerted on the beam, hence, it constitutes a term
of the balance equation, Eq.~(\ref{eq:theory:beam_forces}). The first
derivative term with $\zeta'(z)$ denotes the projection of the tension
force $t$. Due to the clamping, the deflection and its first
derivative must vanish at the end $z=0$. The formula for the
bending moment $m$ is 
\beq{
  m = - E I \zeta'' (z) \;.
  }
The solution $\zeta (z)$ for vanishing $t$ can be obtained as \cite{dubbel_20}
\beq{\label{eq:theory:dubbel}
  \zeta (z) = \frac{f z^2}{12 E I } \left( 3 l  - 2 z \right),
  }
from which we can calculate the elongation
\beq{
   \Delta l = \int_0^l dz \, \sqrt{1+\zeta'^2 (z)} \; -l \approx
   \frac{1}{2} \int_0^l \zeta'^2 \; dz. 
}
It follows from the above equation
\beq{
  t = E S \frac{\Delta l}{l} = E S \epsilon, \label{eq:t_f}
  }
where $S=d^2$ is the beam cross-section area.
The first order solution for $t(f)$ reads as
\beq{ \label{eq:theory:t_firsto}
  t \approx \frac{ l^4 S}{240 E I^2} f^2 \;.
  }

From a computational point of view, a formulation of bending and
stretching in terms of the longitudinal strain $\epsilon$ is more
suitable than using the lateral force $f$. For that, we only need to
replace $m(f)$ by $m(\epsilon)$, which yields 
\beq{
  m(\epsilon) \approx  \frac{f l }{2} = \sqrt{\frac{5}{12}} \frac{E d^4}{l} \sqrt{\epsilon} \label{eq:intro_g},
  }
with
\beq{
  \epsilon = \frac{t}{ES} = \frac{3 l^4}{5 E^2 d^8} f^2 \label{eq:intro_f}.
  }
Using $\epsilon$ as an independent variable enables us to make
comparisons to the simple case of fiber bundle models
\cite{daniels+proc_rsa+1945,sornette_jpa_1989,sornette_jp_1989,pradhan_ijmpb_2003,raul_varint_2002,raul_burst_contdam}
where the elements can have solely stretching deformation.
In the model we represent the interface as an ensemble of parallel
beams connecting the surface of two rigid blocks (see Fig.\
\ref{gr:theory:bendsketch:b}). The beams are
assumed to have identical geometrical extensions (length $l$ and side
length $d$) and linearly elastic behavior characterized by the Young 
modulus $E$. 
In order to capture the failure of the interface in the model, the
beams are assumed to break when their deformation exceeds a certain
threshold value. As it has been shown above, under shear loading of
the interface beams suffer stretching and bending deformation
resulting in two modes of breaking. The two breaking modes can be
considered to be independent or combined in the form of a von Mises type
breaking criterion. The strength of beams is characterized by the two
threshold values of stretching $\epsilon_1$ and bending $\epsilon_2$ a
beam can withstand. The breaking thresholds are assumed to be randomly
distributed variables of the joint probability distribution (PDF)
$p(\epsilon_1,\epsilon_2)$. The 
randomness of the breaking thresholds is supposed to represent the
disorder of the interface material.

 After breaking of a beam the excess load has to be redistributed over
the remaining intact elements. Coupling to the rigid plates ensures that
all the beams have the same deformation giving rise to global load
sharing, {\it i.e.} the load is equally shared by all the elements,
stress concentration in the vicinity of failed beams cannot occur. If
one of the interfaces has a certain compliance, the load
redistribution following breaking of beams becomes localized. This
case has recently been studied for the external load imposed 
perpendicular to the interface~\cite{knudsen_breaksurf_condmat_2004}. 

In the present study we are mainly interested in the macroscopic
response of the interface under shear loading and the process of
progressive failure of interface elements. The global load sharing of
beams enables us to obtain closed analytic results for the
constitutive behavior of the system for both independent and coupled
breaking modes. We examine by computer simulations the
statistics of simultaneously failing elements.
The limiting case of the very localized interaction of interface elements
is explored by computer simulations.

\section{Constitutive behavior}
\label{sec:constit}
Assuming global load sharing for the redistribution of load after the
failure of beams, the most important characteristic quantities of the
interface can be obtained in closed analytic form.

Breaking of the beam is caused by two breaking modes, {\it i.e.}
stretching and bending which can be either independent or coupled by
an empirical breaking criterion. Assuming that the two breaking
modes are independent, a beam breaks if either the longitudinal stress
$t$ or the bending moment $m$ exceeds the corresponding breaking
threshold. Since the longitudinal stress $t$ and the
bending moment $m$ acting on a beam can easily be expressed as 
functions of the longitudinal deformation $\epsilon$, the breaking
conditions can be formulated in a transparent way in terms of
$\epsilon$. To describe the relative importance of the breaking modes,
we assign to each beam two breaking thresholds $\eone^i, \etwo^i$,
$i=1, \ldots , N$, where $N$ denotes the number of beams. The
threshold values $\epsilon_1$ and $\epsilon_2$ are randomly  
distributed according to a joint probability density function
$p(\epsilon_1,\epsilon_2)$ between  lower and upper bounds 
$\epsilon_1^{min}, \epsilon_1^{max}$ and $\epsilon_2^{min},
\epsilon_2^{max}$, respectively. The density function needs to obey
the normalization condition  
\beq{
      \int_{\epsilon_2^{\textrm{min}}}^{\epsilon_2^{\textrm{max}} } d\, \epsilon_2
      \int_{\epsilon_1^{\textrm{min}}}^{\epsilon_1^{\textrm{max}} } d\, \epsilon_1 \;
        p(\epsilon_1, \epsilon_2) =1  .
    }

\subsection{{\it OR} breaking rule}
First, we provide a general formulation of the failure of a bundle of
beams. We allow for two independent breaking modes of a beam that are
functions $f$ and $g$ of the longitudinal deformation
$\epsilon$. Later on this case will be called the {\it OR} breaking 
rule. A single beam breaks if either its stretching or bending
deformation exceed the respective breaking threshold $\epsilon_1$ or
$\epsilon_2$, i.e. failure occurs 
if 
\begin{eqnarray}
  \frac{f(\epsilon)}{\epsilon_1} &\geq& 1 \; \text{or} \label{eq:or_crit_f}  \\*
  \frac{g(\epsilon)}{\epsilon_2} &\geq& 1 ,   \label{eq:or_crit_g}
\end{eqnarray}
where Eqs.~(\ref{eq:or_crit_f},\ref{eq:or_crit_g}) describe the
stretching and bending breaking 
modes, respectively. The functions $f(\epsilon)$ and $g(\epsilon)$ are
called failure functions, for which the only restriction 
is that they be monotonic functions of $\epsilon$. For our specific
case of elastic beams the failure functions can be determined from
Eqs.\ (\ref{eq:t_f},\ref{eq:intro_g}) as
\beq{\label{eq:transfer}
  f(\epsilon) = \epsilon \; , \; g(\epsilon) = a \sqrt{\epsilon} ,
  }
where $a$ is a constant and the value of the Young modulus $E$ is set
to 1.

\begin{figure}[h]
   \includegraphics[clip, 
width=\linewidth]{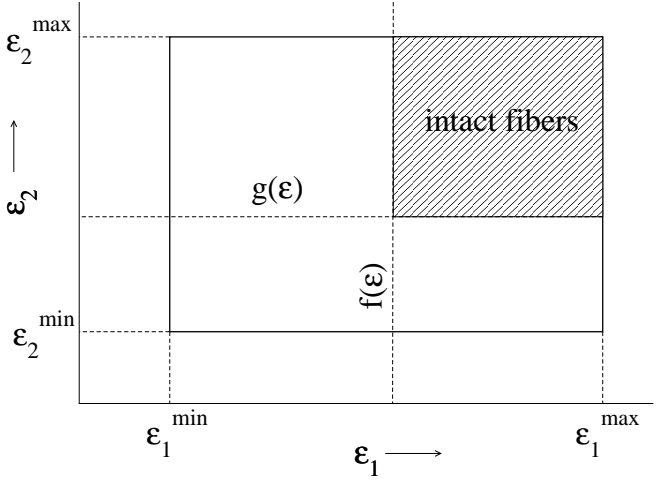}
  \caption{\small Plane of breaking thresholds $(\epsilon_1,
    \epsilon_2)$.  The point of intersection of $f(\epsilon)$ and
    $g(\epsilon)$ determines the fraction of remaining beams.} 
  \label{gr:theory:plane_thresh}
\end{figure}

In the plane of breaking thresholds each point $(\eone, \etwo)$
represents a beam. For each value of $\epsilon$ those beams which
survived the externally imposed deformation are situated in the area 
$f(\epsilon) \leq \epsilon_1 \leq \epsilon_1^{\textrm{max}}$ and
$g(\epsilon) \leq \epsilon_2 \leq \epsilon_2^{\textrm{max}}$, as it is
illustrated in Fig.~\ref{gr:theory:plane_thresh}. 
Hence, the fraction of intact beams $N_{\textrm{intact}}/N$ at a given
value of $\epsilon$ can be obtained by integrating the density
function over the shaded area in Fig.~\ref{gr:theory:plane_thresh} 
\beq{ \label{eq:nintact}
  \frac{N_{\textrm{intact}}}{N} =
  \int_{g(\epsilon)}^{\epsilon_2^{\textrm{max}} } d\, \epsilon_2 
                        \int_{f(\epsilon)}^{\epsilon_1^{\textrm{max}}
                        } d\, \epsilon_1 \; p(\epsilon_1, \epsilon_2) .
    }
Due to the global load sharing,  deformation and stress of the beams
are the same everywhere along the interface. Consequently, the
macroscopic elastic behavior of the system can be obtained by
multiplying the load of a single beam, $\sigma^{(1)} = \epsilon$
($E=1$ is taken), by the fraction of intact elements Eq.~(\ref{eq:nintact})
\beq{ \label{eq:theory:linear}
  \sigma = \epsilon  \int_{g(\epsilon)}^{\epsilon_2^{\textrm{max}} } d\, \epsilon_2
                        \int_{f(\epsilon)}^{\epsilon_1^{\textrm{max}} } d\, \epsilon_1 \;
                        p(\epsilon_1, \epsilon_2) .
    }
Assuming that the breaking thresholds, characterizing the relative
importance of the two breaking modes, are independently distributed, 
the joint PDF can be factorized as 
\beq{
  p(\epsilon_1, \epsilon_2) = p_1(\epsilon_1) \cdot p_2(\epsilon_2) .
  }
Introducing the cumulative distribution functions (CDFs) as
\beq{
  P_1(\epsilon_1) = \int_{\epsilon_1^{\textrm{min}} }^{\epsilon_1} p_1(\epsilon_1 ') \,d \epsilon_1 ', \; \textrm{and}     \;
  P_2(\epsilon_2) = \int_{\epsilon_2^{\textrm{min}} }^{\epsilon_2} p_2(\epsilon_2 ') \,d \epsilon_2 ' \; ,
  }
we can rewrite Eq.\ (\ref{eq:theory:linear}) as
\beqa{ \label{eq:theory:genres}
  \sigma &=& \epsilon \int_{g(\epsilon)}^{\epsilon_2^{\textrm{max}} } d \epsilon_2 \, p_2(\epsilon_2)
                     \int_{f(\epsilon)}^{\epsilon_1^{\textrm{max}} } d \epsilon_1 \, p_1(\epsilon_1) \nonumber \\[.2cm]
          &=& \epsilon [1-P_2(g(\epsilon))][1-P_1(f(\epsilon))].
    }
This is the general formula for the constitutive behavior of a beam
bundle with two breaking modes applying the {\it OR} criterion. 
In the constitutive equation $1-P_1(f(\epsilon))$ and
$1-P_2(g(\epsilon))$ are the fraction of those beams whose threshold
value for bending and stretching is larger than $g(\epsilon)$ and
$f(\epsilon)$, respectively. It follows from the structure of Eq.\
(\ref{eq:theory:genres}) that the existence of two breaking modes
leads to a reduction of the strength of the material, both the critical
stress and strain take smaller values compared to the case of a single
breaking mode applied in simple fiber bundle models \cite{daniels+proc_rsa+1945,sornette_jpa_1989,kloster_pre_1997,raul_varint_2002,raul_burst_contdam,moreno_fbm_avalanche,chakrabarti_fatigue,chakrabarti_phasetrans}.

Considering the special case of two uniform distributions for the
breaking thresholds in the intervals
$[\eonemin , \eonemax]$ and $[\etwomin, \etwomax]$, respectively, we
can derive the specific form of Eq.~(\ref{eq:theory:genres}) by noting
that  
\beq{ \label{eq:uniform}
  p(\eone) = \frac{1}{\eonemax - \eonemin} \; ,\; \; \; p(\etwo) = \frac{1}{\etwomax -\etwomin}
  }
After calculating the cumulative distributions, the final result
follows as
\beq{
  \sigma = \epsilon \frac{[\eonemax - f(\epsilon)][\etwomax - g(\epsilon)]}{[\eonemax - \eonemin][\etwomax - \etwomin]} .
  }
More specifically, if the distributions have equal boundaries $[0,1]$,
and substituting the failure functions $f$ and $g$ from Eq.\ (\ref{eq:transfer}),
the constitutive equation takes the form
\beq{ \label{eq:sp1:const}
  \sigma = \epsilon [1-\epsilon][1 - a\sqrt{\epsilon}] .
  }

\subsection{Von Mises type breaking rule}
\begin{figure}[h]
 \begin{center}
   \includegraphics[clip, width=\linewidth, bb= 50 50 568 568]{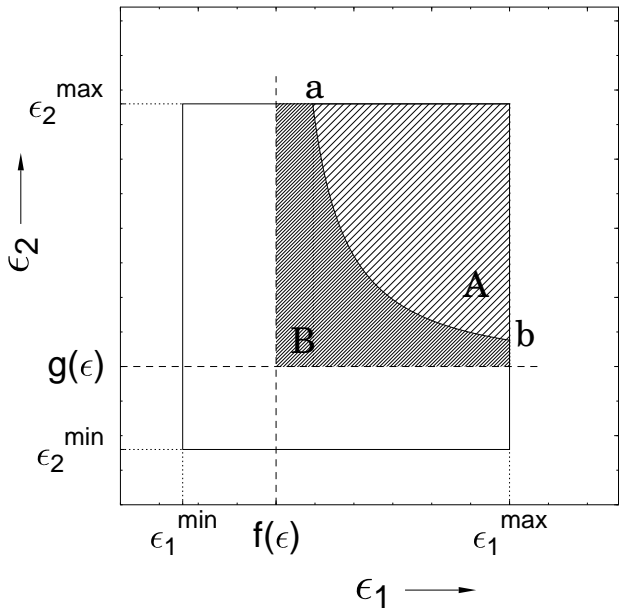}
  \caption{Intact beams in the plane of the failure
    thresholds $\eone, \etwo$ for a given strain $\epsilon$, if
    breaking is determined by the von Mises criterion. The values $a$
    and $b$ are defined as the intersections between the curve of the
    breaking condition Eq.\ (\ref{eq:theory:vm}) and the maximum
    values $\eonemax, \etwomax$, 
    respectively. The shaded region labeled $A$ denotes the 
    intact beams; the shaded region $B$ represents the additionally
    failing beams that would be intact in the case of the
    {\it OR}-criterion.} 
  \label{gr:theory:vm_space}
 \end{center}
\end{figure}

We now address the more complicated case that the two breaking modes
are coupled by a von Mises type breaking criterion: a single beam breaks
if its strain $\epsilon$ fulfills the condition
\cite{hjh_prb_fractdis_1989}
\beq{\label{eq:theory:vm}
  \left( \frac{f(\epsilon)}{\epsilon_1}\right)^2 + \frac{g(\epsilon)}{\epsilon_2} \geq 1  .
  }
This algebraic condition can be geometrically represented as it is 
illustrated in Fig.~\ref{gr:theory:vm_space}. In the plane of
the failure thresholds $\eone, \etwo$, the beams that survive a load
$\epsilon$ are bounded by the maximum values $\eonemax, \etwomax$ and
the hyperbola defined by Eq.~(\ref{eq:theory:vm}).
Calculating the intersection points $a$ and $b$ defined in
Fig.~\ref{gr:theory:vm_space}, which are found to be 
\beqa{
  a &=& f(\epsilon) \left( \frac{\etwomax}{\etwomax - g(\epsilon)} \right)^{1/2} \; \textrm{and} \nonumber \\
  b &=& \frac{g(\epsilon) (\eonemax)^2}{(\eonemax)^2 - f^2 (\epsilon)} \; ,
  }
the fraction of surviving beams can be expressed as
\beq{
  \frac{N_{\text{intact}}}{N} = \int_{a}^{\eonemax} d\epsilon_1 \int_{\tilde{\epsilon}_2 (\eone, \epsilon)}^{\etwomax} d\epsilon_2 \, p(\eone, \etwo) 
  }
with the integration limit
\beq{
  \tilde{\epsilon}_2(\eone, \epsilon) = \frac{\eone ^2 g(\epsilon)}{\eone ^2 - f^2 (\epsilon)} .
  }
The constitutive behavior in this case is therefore given by
\beq{ \label{eq:theory:vm_const}
  \sigma =  \epsilon \int_{a}^{\eonemax} d\eone \int_{\tilde{\epsilon}_2 (\eone,\epsilon )}^{\etwomax} d\epsilon_2 \, p(\eone, \etwo) \; .
  }
We would like to emphasize that assuming independence of the breaking
thresholds the joint distribution factorizes $p(\eone, \etwo) = p_1
(\eone) \cdot p_2 (\etwo) $,  but the integrals in Eq.\
(\ref{eq:theory:vm_const}) over the two variables 
cannot be performed independently. Still, the integral in
Eq.~(\ref{eq:theory:vm_const}) can be evaluated analytically for a
broad class of disorder distributions. As an example, we again
consider two homogeneous distributions Eq.\ (\ref{eq:uniform})
over the interval $[0, 1]$ along with the failure functions Eq.\
(\ref{eq:transfer}). Setting the Young modulus and the parameter $E=1=a$, the integrals
yield 
\beqa{ \label{eq:sp_vm_res}
   \sigma &=&   \epsilon \cdot  \frac{1}{2}  \left[  \left( 2 - 2\,\sqrt{\epsilon} + \epsilon^{\frac{3}{2}} \,\log \frac{ 1 + \epsilon}{1 - \epsilon} \right) \right.  \nonumber \\*
          &-&  \left. \epsilon^{3/2} \left( 2 \sqrt{\frac{1 - \sqrt{\epsilon}}{\epsilon}} + \log \frac{1+\sqrt{1-\sqrt{\epsilon}}}{1- \sqrt{1-\sqrt{\epsilon}}} \right) \right] .
 }

Even for the simplest case of uniformly distributed breaking
thresholds, the constitutive equation takes a rather complex form. It
is important to note that the coupling of the two breaking modes gives
rise to a higher amount of broken beams compared to the {\it OR}
criterion. In Fig.\ \ref{gr:theory:vm_space} the beams which break due
to the coupling of the two breaking modes fall in the area labeled by
$B$. 
 
\section{Computer simulations} 
\label{sec:sim}
\begin{figure}[h]
 \begin{center}
  \includegraphics[clip, width=\linewidth]{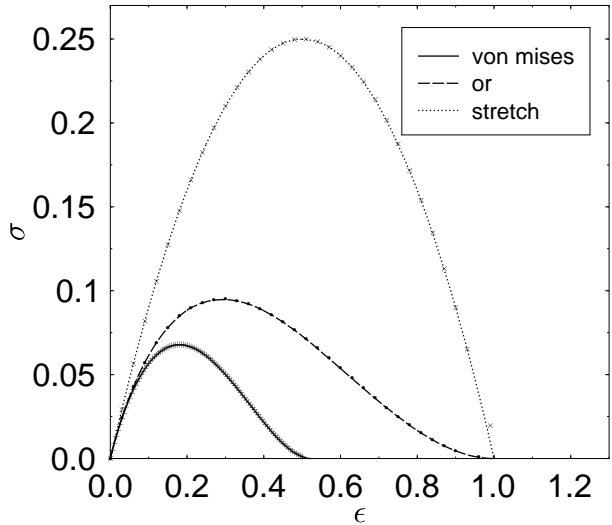}
  \caption{\small Constitutive behavior of a bundle of beams with two
    breaking modes in a strain-controlled simulation of $N=4 \cdot
    10^5$ beams, under the $OR$ (dashed line), {\it von Mises type} (solid
    line), and a pure stretching breaking criterion (dotted line). The
    random failure thresholds for the breaking modes
    of each beam are sampled from uniform distribution between
    $[0,1]$. The points marked with `$\cdot$', `$+$' and `$\times$'
    denote the respective theoretical results,
    Eqs.~(\ref{eq:sp1:const}, \ref{eq:sp_vm_res}), and $\sigma =
    \epsilon(1- \epsilon)$ for the pure stretching case. The
    constants $E$ and $a$ are set to unity here.} 
  \label{gr:theory:beam_const} 
 \end{center}
\end{figure}
In order to determine the behavior of the system for complicated
disorder distributions and explore the microscopic failure process of
the sheared interface, it is necessary to work out a computer
simulation technique. In the model we consider an ensemble of $N$ beams
arranged on a square lattice. Two breaking thresholds
$\epsilon_1^i, \epsilon_2^i$ are assigned  to each
beam $i$  ($i=1,\ldots ,N$) of the bundle from the joint probability
distribution 
$p(\epsilon_1,\epsilon_2)$. For the $OR$ breaking rule, the failure of
a beam is caused either by stretching or bending depending on which
one of the conditions Eqs.\ (\ref{eq:or_crit_f},\ref{eq:or_crit_g}) is
fulfilled at a lower value of the external load. This way an
effective breaking threshold $\epsilon_c^i$ can be defined for the beams
as 
\beq{\label{eq:effective}
\epsilon_c^i = \min(f^{-1} (\eone^i), g^{-1} (\etwo^i)), \ \ \ \ \ i=1,
\ldots , N,
}
where $f^{-1}$ and $g^{-1}$ denote the inverse of $f, g$,
respectively. A beam $i$ breaks during the loading process of the
interface when the load on it exceeds its effective breaking threshold
$\epsilon_c^i$. 
For the case of the von Mises type breaking criterion Eq.\
(\ref{eq:theory:vm}), the effective breaking threshold $\epsilon_c^i$
of beam $i$ can be obtained as the solution of the algebraic equation 
\beq{\label{eq:theory:vm_fgsimple}
  \left( \frac{f(\epsilon_c^i)}{\epsilon_1^i}\right)^2 +
  \frac{g(\epsilon_c^i)}{\epsilon_2^i} = 1, \ \ \ \ \  i=1,
\ldots , N. 
  }
Although for the specific case of the functions $f,g$ given by Eqs.\
(\ref{eq:or_crit_f},\ref{eq:or_crit_g}) the above equation 
can be converted to a 4th order
polynomial and solved analytically, this solution turns out to be
impractical, especially since the numerical evaluation of the solution
is too slow. We therefore solve Eq.~(\ref{eq:theory:vm_fgsimple})
numerically by means of a modified Newton root finding scheme, where
we make use of the fact that the solution has the lower bound 0.  

\begin{figure}
 \begin{center}
  \includegraphics[clip, width=\linewidth,bb= 50 50 568 568]{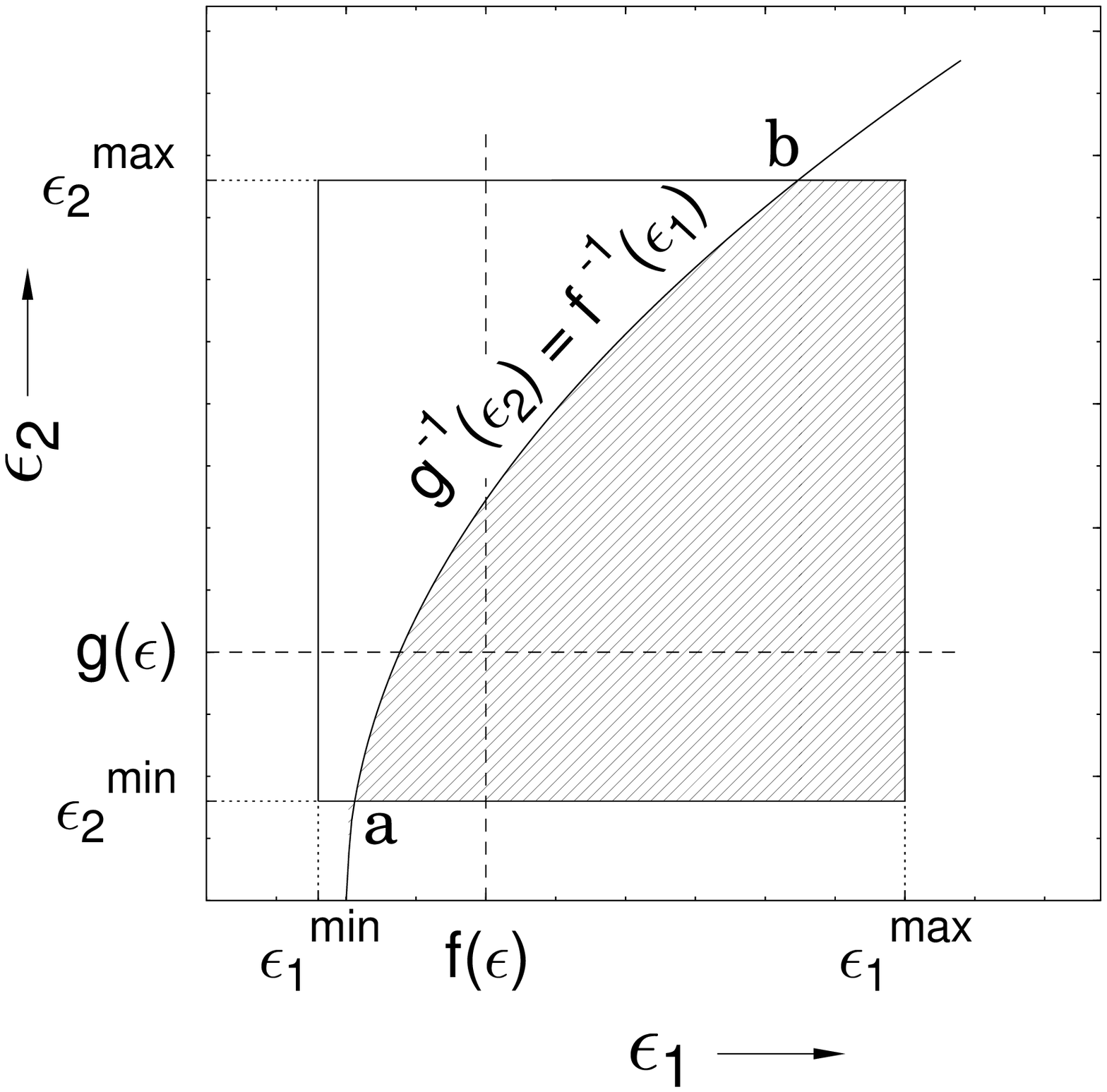}
  \caption{\small The beams that break due to mode $g$ fall in the shaded
    region. The labels $a$ and $b$ mark the abscissae of the
    intersection points of the curve $g^{-1} (\etwo) = f^{-1}
    (\eone)$ with the lines $\etwo = \etwomin$ and $\etwo=\etwomax$,
    respectively.}
  \label{gr:fracbroken1}
 \end{center}
\end{figure}
In the case of global load sharing, the load and deformation of beams
is everywhere the same along the interface, which implies that beams
break in the increasing order of their effective breaking thresholds.
In the simulation, after determining $\epsilon_c^i$ for each beam,
they are sorted in increasing order. Quasi-static loading of the
beam bundle is performed by 
increasing the external load to break only a single
element. Due to the subsequent load redistribution on the intact
beams, the failure of a beam may trigger an avalanche of breaking
beams. This process has to be iterated until the avalanche stops, or
it leads to catastrophic failure at the critical stress and strain.  
Under strain controlled loading conditions, however, the load of the
beams is always determined by their deformation so that there is no
load redistribution and avalanche activity. 

In Fig.\ \ref{gr:theory:beam_const} the analytic results of Sec.\
\ref{sec:constit} on the constitutive behavior Eqs.\
(\ref{eq:sp1:const}, \ref{eq:sp_vm_res}) are
compared to the corresponding results of computer simulations. As a
reference, we also plotted the constitutive behavior of a bundle of
fibers where the fibers fail solely due to simple stretching
\cite{daniels+proc_rsa+1945,sornette_jpa_1989,kloster_pre_1997,raul_varint_2002,raul_burst_contdam,moreno_fbm_avalanche,chakrabarti_fatigue,chakrabarti_phasetrans}.
It can be seen in the figure that the 
simulation results are in perfect agreement with the analytical
predictions. It is important to note that the presence of two breaking
modes substantially reduces the critical stress $\sigma_c$ and strain
$\epsilon_c$ ($\sigma$ and $\epsilon$ value of the maximum of the
constitutive curves) with respect to the case when failure of elements
occurs solely under stretching. Since one of the failure functions
$g(\epsilon)$ is non-linear, the shape of the constitutive curve
$\sigma(\epsilon)$ also changes, especially in the post-peak
regime. The coupling of the two breaking modes 
in the form of the von Mises criterion gives rise to further reduction
of the strength of the interface.
\begin{figure}
 \begin{center}
  \includegraphics[clip, width=\linewidth,bb=0 0 280
  280]{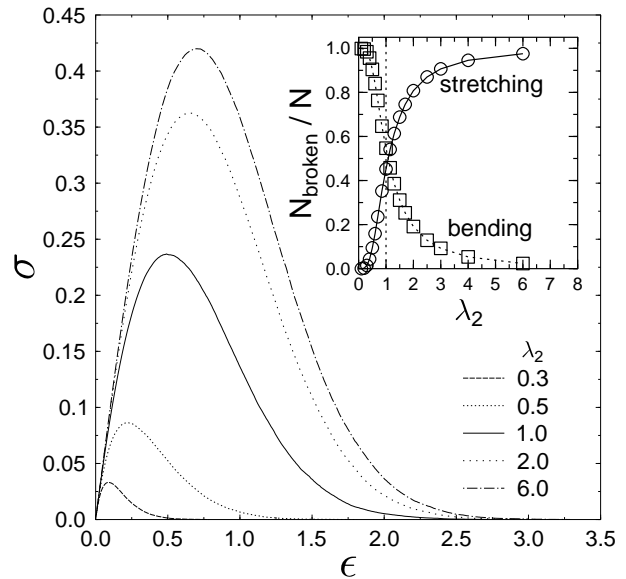}  
  \caption{\small Constitutive behavior of a bundle of $N=90000$
    beams using the {\it OR} criterion. The parameter values
    $\lambda_1=1.0$ (stretching), $m_1 
    = m_2 = 2$ were fixed, while $\lambda_2$ corresponding to the
    bending mode was shifted.
    Inset: Fraction of beams breaking by stretching and bending
    as a function of $\lambda_2$.
}
  \label{gr:gls_or_shiftl_fff}
 \end{center}
\end{figure}

\section{Progressive failure of the interface}
During the quasi-static loading process of an interface,
avalanches of simultaneously failing beams occur. Inside an
avalanche, however, the beams can break under different breaking
modes when the {\it OR} criterion is considered, or the breaking can
be dominated by one of the breaking modes in the coupled case of the
von Mises type criterion. Hence, it is an important question how the fraction of beams breaking
due to a specific breaking mode (stretching or bending) varies during
the course of loading of the interface.

For the {\it OR} criterion, those beams break, for instance, under
bending, {\it i.e.} under mode 
$g$ defined by Eq.\ (\ref{eq:or_crit_g}), whose effective breaking
threshold $\epsilon_c^i$ is determined by $g^{-1}(\epsilon_2^i)$ in Eq.\
(\ref{eq:effective}) so that  the inequality holds
\beq{ \label{eq:frac_mode_g}
  g^{-1} (\etwo^i) < f^{-1} (\eone^i).
  }
\begin{figure}
 \begin{center}
  \includegraphics[clip, width=\linewidth,bb=0 0 280 280]{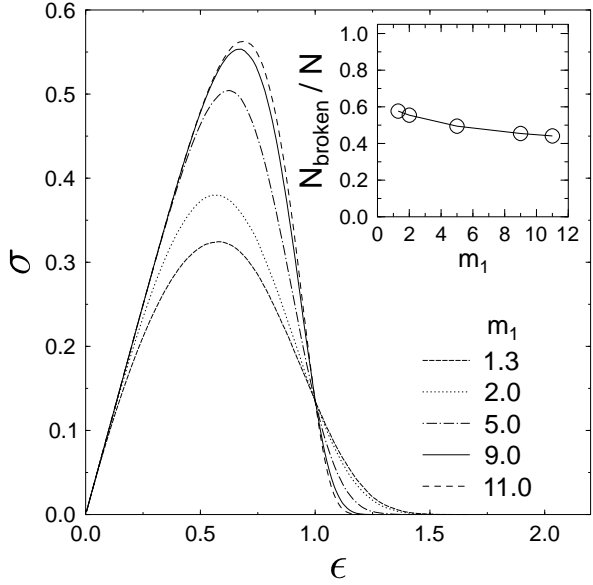}
  \caption{\small Constitutive behavior for different values of the
    shape parameter $m_1$ of stretching. Strain controlled simulation
    of $N=90000$ beams with  failure due to the {\it OR}-criterion,
    fixing the parameters  $\lambda_1=\lambda_2=1.0$  and  $m_2 =
    2$. Inset: total fraction of beams broken under mode $g$ during the
    course of loading.
  } 
  \label{gr:gls_or_shiftm_const}
 \end{center}
\end{figure}
In the plane of breaking thresholds $\{\epsilon_1, \epsilon_2 \}$ the
region of beams which fulfill the above condition is indicated by
shading in  Fig.~\ref{gr:fracbroken1}.
The fraction of beams $ {\mathcal{B}}_g(\epsilon)$ breaking under mode
$g$ up to the macroscopically imposed deformation $\epsilon$ can be
obtained by integrating the probability distribution $p(\epsilon_1,
\epsilon_2)$ over the shaded area in Fig.\ \ref{gr:fracbroken1}.
Taking into account the fact that the intersection points $a,b$
defined in Fig.~\ref{gr:fracbroken1} may in general lie outside the
rectangle $(\eonemin, \eonemax, \etwomax, \etwomin)$ and adjusting the
integral limits accordingly, we arrive at the following formula for
the fraction of fibers breaking under mode $g$ as a function of the
deformation $\epsilon$ 
\beqa{
&&  {\mathcal{B}}_g (\epsilon) = \int \limits_{\max
  (\eonemin,a)}^{\min (f(\epsilon), b)} d \eone \int
\limits_{\etwomin}^{g(f^{-1}(\eone))} d \etwo \>  p(\eone, \etwo)
\nonumber \\* 
&&                  + \int \limits_{\min
  (f(\epsilon),b)}^{f(\epsilon)} d \eone \int
\limits_{\etwomin}^{\etwomax} d \etwo \>  p(\eone, \etwo) \nonumber \\ 
&&                  + \int \limits_{f(\epsilon)}^{\eonemax} d \eone \int
                  \limits_{\etwomin}^{g(\epsilon)} d \etwo \>
                  p(\eone, \etwo). \label{eq:bg_epsilon} 
  }
It should be noted that the second integral vanishes unless $b < \eonemax$. 
The total fraction of beams breaking under mode $g$ during the entire
course of the loading can be obtained by substituting
$\epsilon=\epsilon^{max}$ in the above formulas, where
$\epsilon^{max}$ denotes the deformation at the breaking of the last
beam. 

In order to study the effect of the disorder distribution
$p(\epsilon_1,\epsilon_2)$ of beams on the relative importance
of the two breaking modes and on the progressive failure of the
interface, we considered independently distributed breaking thresholds
$\epsilon_1, \epsilon_2$ both with a Weibull distribution
\begin{eqnarray}
\label{eq:weibull}
p_b(\epsilon_b) =
{\displaystyle
  \frac{m_b}{\lambda_b}\left(\frac{\epsilon_b}{\lambda_b}\right)^{m_b-1} 
\exp{\left[-\left(\frac{\epsilon_b}{\lambda_b}   \right)^{m_b}\right] }
 },
\end{eqnarray}
where index $b$ can take values $1$ and $2$. The exponents $m_1, m_2$
determine the amount of disorder in the system for stretching and
bending, respectively, {\it i.e.} the width
of the distributions Eq.\ (\ref{eq:weibull}), while the values of
$\lambda_1, \lambda_2$ set the average strength of beams for the two
breaking modes. Computer simulations were performed in the framework
of global load sharing by setting equal values for the shape
parameters $m_1=m_2$ and fixing the value of $\lambda_1=1$ of the
stretching mode, while varying $\lambda_2$ of the bending mode.      

The total fraction of beams breaking by stretching and bending using the
{\it OR} breaking rule is presented in Fig.\
\ref{gr:gls_or_shiftl_fff}. Increasing $\lambda_2$ of the bending
mode, the beams become more resistant against bending so that the
stretching mode starts to dominate the breaking of beams, which is
indicated by the increasing fraction of stretching failure in the figure. In the
limiting case of $\lambda_2 >> \lambda_1$ the beams solely break under
stretching. Decreasing $\lambda_2$ has the opposite effect, more and
more beams fail due to bending, while the fraction of beams breaking
by the stretching mode tends to zero.
It is interesting to note that varying the relative importance of the
two failure modes gives also rise to a change of the macroscopic
constitutive behavior of the system. Fig.\
\ref{gr:gls_or_shiftl_fff} illustrates that shifting the strength
distributions of beams the functional form of the constitutive
behavior remains the same, however, the value of the critical stress
and strain vary in a relatively broad range.
\begin{figure}
 \begin{center}
  \includegraphics[clip, width=\linewidth,bb=90 350 430
  640]{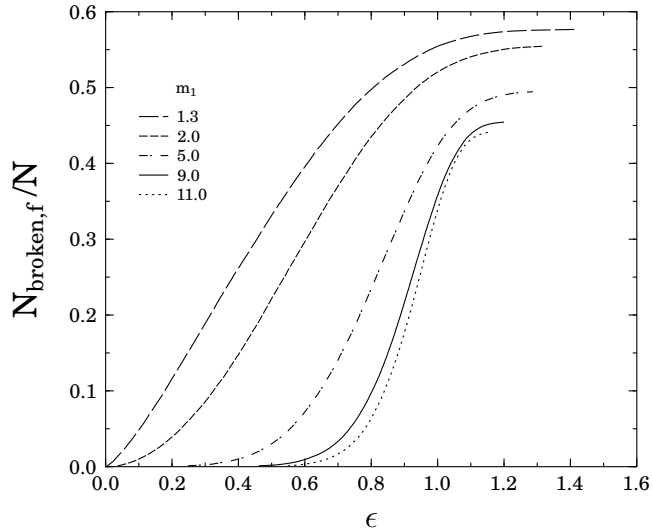} 
  \caption{\small Fraction of fibers broken by the stretching  mode as
    a function of $\epsilon$ for different values of the corresponding
    shape parameter $m_1$. Strain controlled simulation with failure
    due to the {\it OR}-criterion, $N=90000$,
    $\lambda_1=\lambda_2=1.0$, $m_2 = 9$.  } 
  \label{gr:gls_or_shiftm_ff}
 \end{center}
\end{figure}

The same analysis can also be performed by fixing the values
$\lambda_1$ and $\lambda_2$ and changing the relative width of the two
distributions by varying one of the Weibull shape parameters $m$.
We find it convenient to shift
$m_1$, the shape parameter of the stretching mode instead of $m_2$. It
can be observed in Fig.~\ref{gr:gls_or_shiftm_const}
that for this choice of the scale parameters
$\lambda$, the value of the critical strain hardly changes, however
the critical stress nearly doubles as compared to
Fig.~\ref{gr:gls_or_shiftl_fff}. 

Although the effect on the final fraction of beams broken by each
mode, see inset of Fig.~\ref{gr:gls_or_shiftm_const}, is not as
pronounced as for 
shifting $\lambda$, we should also consider the fraction of fibers
broken up to a value of $\epsilon$ during the loading process
(Fig.~\ref{gr:gls_or_shiftm_ff}). It 
should be noted that the end points of the respective curves in
Fig.~\ref{gr:gls_or_shiftm_ff} are just the final fraction numbers in
Fig.~\ref{gr:gls_or_shiftm_const}, but the curves show a strong spread
for intermediate values of $\epsilon$. This demonstrates that changing
the amount of disorder in the breaking thresholds strongly influences
the process of damaging of the interface.

We apply the methods outlined in the previous paragraphs to the von
Mises case. Obviously, Eq.~(\ref{eq:theory:vm}) does not allow for a
strict separation of the two modes. However, the breaking of a beam at
a certain value $\epsilon_c$ is dominated by stretching if 
\beq{
\left( \frac{f(\epsilon)}{\eone} \right)^2 > \frac{g(\epsilon)}{\etwo}. 
}

\begin{figure}
 \begin{center}
  \includegraphics[clip, width=\linewidth,bb=0 0 280 280]{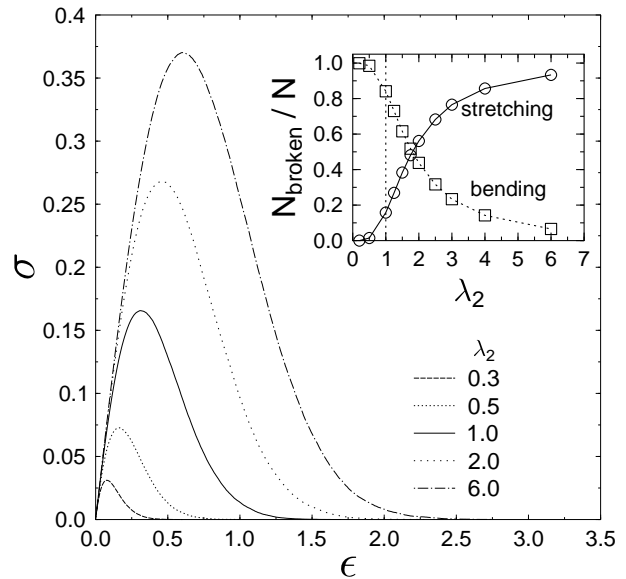} 
  \caption{\small Constitutive behavior for different values of the
    bending scale parameter $\lambda_2$. Strain controlled simulation
    with the von Mises criterion , $N=90000$, $\lambda_1=1.0$, $m_1=
    m_2 = 2$.  The inset presents the fraction of beams whose failure
    was dominated by the stretching or bending mode.} 
  \label{gr:gls_vm_shiftl_const}
 \end{center}
\end{figure}

With the previous prescriptions for the failure
functions Eqs.\ (\ref{eq:transfer}), we again find a
massive influence on the constitutive behavior and the final number of broken
beams, see Fig.~\ref{gr:gls_vm_shiftl_const}. The inset of
Fig.~\ref{gr:gls_vm_shiftl_const} demonstrates that a crossover
between stretching and bending preponderance occurs also in the von Mises case.

\section{Avalanche statistics}
The stress controlled loading of the glued interface is accompanied by
avalanches of simultaneously failing elements. The avalanche activity
can be characterized by the distribution $D(\Delta)$ of burst sizes
$\Delta$ defined as the number of beam breakings triggered by the
failure of a single beam. 
In the framework of simple fiber bundle models, it has been shown
analytically that global load sharing gives rise to a power law
distribution of avalanche sizes for a very broad class of disorder
distributions of materials strength
\cite{hemmer_distburst_jam_1992,kloster_pre_1997}
\beq{
  D(\Delta) \propto \Delta^{-\delta}
}
with an universal exponent $\delta=5/2$. 

\begin{figure}
 \begin{center}
  \includegraphics[clip, width=\linewidth]{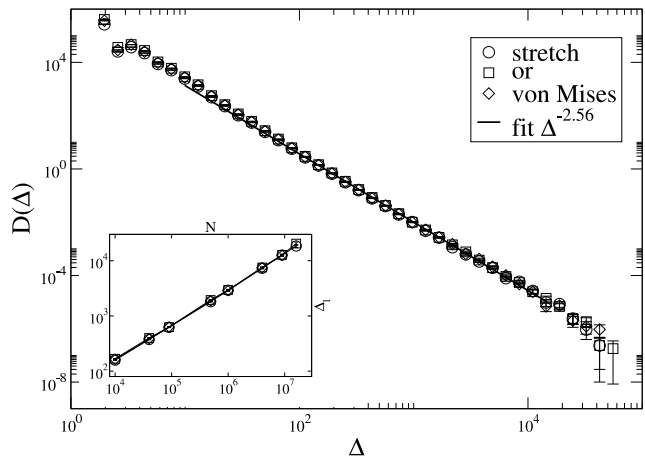}
  \caption{\small Avalanche size distribution $D(\Delta)$ for pure
    stretching of a fiber bundle, and the two beam breaking conditions
    for system sizes $N=16 \cdot 10^6$, averaged over 100 runs. A fit
    with the best result $D \propto \Delta^{-2.56}$ over almost four
    decades is provided. The inset shows the dependency of the largest
    avalanche $\Delta_l$ on the system size for the three
    cases. Again, no difference is found. } 
  \label{gr:aval1}
 \end{center}
\end{figure}

In the previous sections we have shown that in our model the interplay
of the two breaking modes results in a complex failure mechanism on the
microscopic level, which is strongly affected by the distributions of
the breaking thresholds. 
In order to investigate the bursts of breaking beams
we performed stress controlled  simulations on large systems
($N=10^4 \dots 16 \cdot 10^6$) with both the {\it OR} and von Mises
type breaking criterion. In Figure \ref{gr:aval1} the simulation
results are compared to the avalanche size distribution of a simple
fiber bundle model where failure occurs solely due to stretching
\cite{hemmer_distburst_jam_1992,kloster_pre_1997,raul_varint_2002,raul_burst_contdam,moreno_fbm_avalanche}.
In all the cases the avalanche size distributions can be fitted by a
power law over three orders of magnitude. The best fit exponent of
$\delta = 2.56 \pm 0.08$ was derived from the system of size $N=16
\cdot 10^6$ beams, with an average taken over 100 samples. 
The size of the largest avalanche in the inset of Fig.\ \ref{gr:aval1}
proved to be proportional to the system size.
It can be concluded that the beam model belongs to the same
universality class as the fiber bundle model
\cite{hemmer_distburst_jam_1992,kloster_pre_1997,raul_varint_2002,raul_burst_contdam,moreno_fbm_avalanche}.

\section{Local load sharing}
\begin{figure}
   \includegraphics[clip, width=\linewidth, bb= 15 15 700 660
   ]{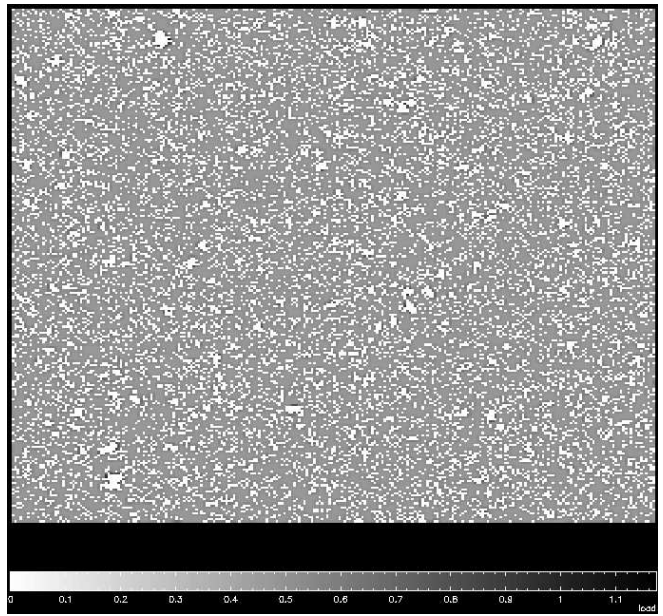} 
   \caption{\small Snapshot of a LLS system at the last stable
     configuration. The color coding represents the load per beam,
     with broken beams carrying a vanishing load. The  system size is
     $L=100$.  } 
  \label{gr:lls_snapshot}
\end{figure}

During the failure of interfaces, stress localization is known to
occur in the vicinity of failed regions, which results in correlated
growth and coalescence of cracks. In our model this effect can be
captured by localized interaction of the interface elements, which
naturally occurs when the two solid blocks are not perfectly rigid
\cite{batrouni_intfail_pre_2002}. For simplicity, in our model solely
the extremal case of very localized interactions is considered, {\it
  i.e.} after breaking of a beam in the square lattice, the load is
redistributed equally on its nearest and next-nearest intact
neighbors. This localized load sharing (LLS) results in growing failed
regions (cracks) with high stress concentration along their perimeter
\cite{batrouni_intfail_pre_2002,raul_varint_2002,hansen_distburst_local_1994}.
Figure \ref{gr:lls_snapshot} shows the last 
stable configuration of a beam lattice preceding global failure,
which was obtained using the {\it OR} criterion for beam breaking. Due
to the stress concentration around cracks, the onset of a catastrophic
avalanche occurs at lower external loads making the macroscopic
response of the interface more brittle compared to the case of global load
sharing \cite{batrouni_intfail_pre_2002,raul_varint_2002,hansen_distburst_local_1994}.

\begin{figure}
   \includegraphics[clip, width=\linewidth, bb= 10 10 650 520
   ]{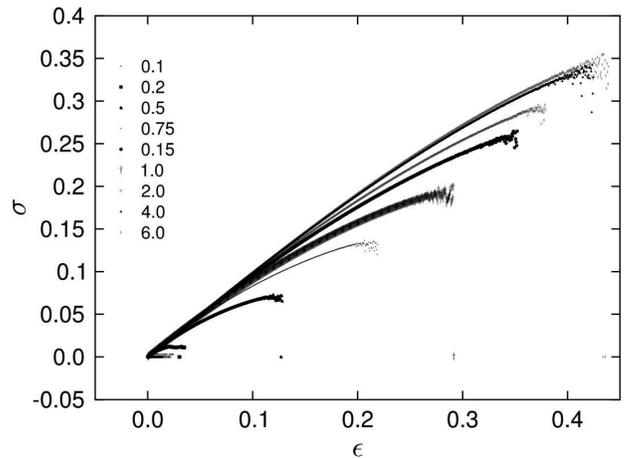} 
   \caption{\small Constitutive curves in the LLS case, shifting
     $\lambda_2$ and keeping the parameters $\lambda_1=1.0$ and $m_1=m_2=2$
     fixed. Stress controlled simulation of  $N=10000$ fibers
     averaged over 300 runs. } 
  \label{gr:lls_or_const_l2}
\end{figure}

As for global load sharing, we shift the relative importance of the
two breaking modes by changing their threshold distributions, and
record the influence on micro- and macroscopic 
system properties. We consider here the {\it OR}-criterion, and use
two Weibull distributions with parameters $\lambda_1, \lambda_2$ and
$m_1, m_2$, where the indices 1 and 2 denote the stretching and
bending mode, 
respectively. Varying $\lambda_2$ for a fixed $\lambda_1$, we find a
considerable influence on the constitutive properties, as
Fig.~\ref{gr:lls_or_const_l2} illustrates.  

\begin{figure}
 \begin{center}
  \includegraphics[clip,
  width=\linewidth]{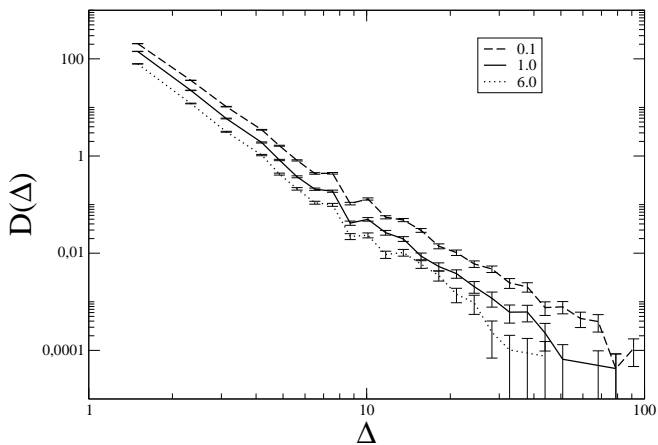} 
  \caption{\small Distribution of avalanche sizes for LLS for three
    values of  $\lambda_2$, with $\lambda_1=1$ and $m_1 = m_2 = 2$
    fixed. Simulations were performed using the {\it OR} criterion for
    a bundle of 10000 beams averaged over 300 runs.  }  
  \label{gr:lls_or_shiftl_aval}
 \end{center}
\end{figure}

We investigated also  the distribution of avalanche sizes for LLS,
Fig.~\ref{gr:lls_or_shiftl_aval}, where we vary the  scale parameter
$\lambda_2$ of the bending mode $g$. We find merely a shifting to
different amplitudes, but no considerable effect on the shape of the
distribution function, which is similar to the one reported in
\cite{raul_varint_2002}. In 
comparison to the GLS case, we should note that large avalanches cannot
occur, and the functional form of the curves can be approximated by a
power law with an exponent higher than for GLS in agreement with
Refs.\
\cite{hemmer_distburst_jam_1992,kloster_pre_1997,hansen_distburst_local_1994}.

\section{Concluding remarks}
Fiber bundle models have been applied to describe various
aspects of the failure of heterogeneous interfaces. However, fibers
can sustain solely elongation, and hence cannot account for
more complex deformation states of interface elements, which naturally
occurs under shear loading.
We constructed a novel type of model for the shear
failure of the glued interface of two solid blocks. 
In the model the interface is discretized in terms of elastic beams
which experience stretching and bending deformation under
shear. Breaking of a beam can be caused by both deformations resulting
in two failure modes. The mechanical strength of beam elements is
characterized by the two threshold values of stretching and bending
the beam can withstand. The beams are assumed to have identical
elastic properties, the heterogeneous microstructure is represented by
the disorder distribution of the breaking thresholds. In the model we
assume that the two solid blocks are perfectly rigid which results in
a global load redistribution over the intact beams following the breaking
events. 

We presented a detailed study of the macroscopic response 
and the progressive damaging of the interface under quasistatic
loading. Making use of the global load sharing of intact beams, we
obtained the analytic solution of the model for the constitutive behavior
and the amount of damage during the course of loading.  
In order to explore the microscopic process of damaging we worked out
an efficient simulation technique which enables us to study large
systems. We demonstrated that the disorder distribution and the relative
importance of the two failure modes have a substantial effect on both
the microscopic damage process and the macroscopic constitutive
behavior of the interface. Varying its parameters, the model provides
a broad spectrum of material behaviors. 
Simulations showed that the failure of the interface proceeds in bursts
of simultaneously breaking beams. The distribution of burst sizes
follows power law behavior with an exponent equal to the one of simple
fiber bundles. Under stress controlled loading conditions, the
macroscopic failure of the interface occurs analogously to phase
transitions, where our beam model proved to be in the same
universality class as the equal load sharing fiber bundle model
\cite{andersen_tricrup_prl_1997,raul_varint_2002,pradhan_ijmpb_2003}.  
We showed that the localized interaction of beams leads
to a more brittle behavior of the interface, which implies a more
abrupt transition at the critical load. 

Beam models have been successfully applied to study the fracture of
cohesive frictional materials where cracks usually form along the
grain-grain interface. Beam elements proved to give a satisfactory
description of the interfacial failure of grains
and the emerging micro- and
macro-behavior of materials \cite{HJH_cont_and_discont_2000}. 
Our beam model presented here provides a more realistic description of
the interface of macroscopic solid bodies than the simple fiber bundle
model and is applicable to more complex loading
situations. Experiments on the shear failure of glued interfaces are
rather limited, especially on the microscopic mechanism of the
progressive damage, which hinders the direct comparison of our
theoretical results to experimental findings. Our work in this direction
is in progress.

\begin{acknowledgments}
We wish to thank F.\ Wittel for useful discussions. This work was
supported by the Collaborative Research Center SFB381. F.\ Kun 
acknowledges financial support of the Research Contracts 
NKFP-3A/043/04, OTKA M041537, T049209 and of the Gy\"orgy B\'ek\'esi
Foundation of the Hungarian Academy of Sciences. 
\end{acknowledgments}

\bibliography{mybib}

\begin{thebibliography}{25}
\expandafter\ifx\csname natexlab\endcsname\relax\def\natexlab#1{#1}\fi
\expandafter\ifx\csname bibnamefont\endcsname\relax
  \def\bibnamefont#1{#1}\fi
\expandafter\ifx\csname bibfnamefont\endcsname\relax
  \def\bibfnamefont#1{#1}\fi
\expandafter\ifx\csname citenamefont\endcsname\relax
  \def\citenamefont#1{#1}\fi
\expandafter\ifx\csname url\endcsname\relax
  \def\url#1{\texttt{#1}}\fi
\expandafter\ifx\csname urlprefix\endcsname\relax\def\urlprefix{URL }\fi
\providecommand{\bibinfo}[2]{#2}
\providecommand{\eprint}[2][]{\url{#2}}

\bibitem[{\citenamefont{Cahn et~al.}(1993)\citenamefont{Cahn, Haasen, and
  Kramer}}]{cahn}
\bibinfo{author}{\bibfnamefont{R.~W.} \bibnamefont{Cahn}},
  \bibinfo{author}{\bibfnamefont{P.}~\bibnamefont{Haasen}}, \bibnamefont{and}
  \bibinfo{author}{\bibfnamefont{E.~J.} \bibnamefont{Kramer}},
  \emph{\bibinfo{title}{Structure and Properties of Composites}}
  (\bibinfo{publisher}{VCH-{V}erlag Weinheim, Germany}, \bibinfo{year}{1993}).

\bibitem[{\citenamefont{Herrmann and Roux}(1990)}]{hh_smfdm}
\bibinfo{editor}{\bibfnamefont{H.~J.} \bibnamefont{Herrmann}} \bibnamefont{and}
  \bibinfo{editor}{\bibfnamefont{S.}~\bibnamefont{Roux}}, eds.,
  \emph{\bibinfo{title}{Statistical models for the fracture of disordered
  media}}, Random materials and processes (\bibinfo{publisher}{Elsevier},
  \bibinfo{address}{Amsterdam}, \bibinfo{year}{1990}).

\bibitem[{\citenamefont{Vermeer et~al.}(2000)}]{hh_cdm}
\bibinfo{editor}{\bibfnamefont{P.}~\bibnamefont{Vermeer}} \bibnamefont{et~al.},
  eds., \emph{\bibinfo{title}{Continuous and {D}iscontinuous {M}odeling of
  {C}ohesive-{F}rictional {M}aterials}}, Lecture {N}otes in {P}hysics
  (\bibinfo{publisher}{Springer}, \bibinfo{address}{New {Y}ork and others},
  \bibinfo{year}{2000}).

\bibitem[{\citenamefont{Daniels}(1945)}]{daniels+proc_rsa+1945}
\bibinfo{author}{\bibfnamefont{H.~E.} \bibnamefont{Daniels}},
  \bibinfo{journal}{Proc. {R.} {S}oc {L}ondon {A}}
  \textbf{\bibinfo{volume}{183}}, \bibinfo{pages}{405} (\bibinfo{year}{1945}).

\bibitem[{\citenamefont{Sornette}(1989{\natexlab{a}})}]{sornette_jpa_1989}
\bibinfo{author}{\bibfnamefont{D.}~\bibnamefont{Sornette}},
  \bibinfo{journal}{J. Phys. A} \textbf{\bibinfo{volume}{22}},
  \bibinfo{pages}{L243} (\bibinfo{year}{1989}{\natexlab{a}}).

\bibitem[{\citenamefont{Kloster et~al.}(1997)\citenamefont{Kloster, Hansen, and
  Hemmer}}]{kloster_pre_1997}
\bibinfo{author}{\bibfnamefont{M.}~\bibnamefont{Kloster}},
  \bibinfo{author}{\bibfnamefont{A.}~\bibnamefont{Hansen}}, \bibnamefont{and}
  \bibinfo{author}{\bibfnamefont{P.~C.} \bibnamefont{Hemmer}},
  \bibinfo{journal}{Phys. {R}ev. {E}} \textbf{\bibinfo{volume}{56}},
  \bibinfo{pages}{2615} (\bibinfo{year}{1997}).

\bibitem[{\citenamefont{Hidalgo et~al.}(2002)\citenamefont{Hidalgo, Moreno,
  Kun, and Herrmann}}]{raul_varint_2002}
\bibinfo{author}{\bibfnamefont{R.~C.} \bibnamefont{Hidalgo}},
  \bibinfo{author}{\bibfnamefont{Y.}~\bibnamefont{Moreno}},
  \bibinfo{author}{\bibfnamefont{F.}~\bibnamefont{Kun}}, \bibnamefont{and}
  \bibinfo{author}{\bibfnamefont{H.~J.} \bibnamefont{Herrmann}},
  \bibinfo{journal}{Phys. Rev. E} \textbf{\bibinfo{volume}{65}},
  \bibinfo{pages}{046148} (\bibinfo{year}{2002}).

\bibitem[{\citenamefont{Hidalgo et~al.}(2001)\citenamefont{Hidalgo, Kun, and
  Herrmann}}]{raul_burst_contdam}
\bibinfo{author}{\bibfnamefont{R.~C.} \bibnamefont{Hidalgo}},
  \bibinfo{author}{\bibfnamefont{F.}~\bibnamefont{Kun}}, \bibnamefont{and}
  \bibinfo{author}{\bibfnamefont{H.~J.} \bibnamefont{Herrmann}},
  \bibinfo{journal}{Phys. Rev. E} \textbf{\bibinfo{volume}{64}},
  \bibinfo{pages}{066122} (\bibinfo{year}{2001}).

\bibitem[{\citenamefont{Moreno et~al.}(2000)\citenamefont{Moreno, Gomez, and
  Pacheco}}]{moreno_fbm_avalanche}
\bibinfo{author}{\bibfnamefont{Y.}~\bibnamefont{Moreno}},
  \bibinfo{author}{\bibfnamefont{J.~B.} \bibnamefont{Gomez}}, \bibnamefont{and}
  \bibinfo{author}{\bibfnamefont{A.~F.} \bibnamefont{Pacheco}},
  \bibinfo{journal}{Phys. Rev. Lett.} \textbf{\bibinfo{volume}{85}},
  \bibinfo{pages}{2865} (\bibinfo{year}{2000}).

\bibitem[{\citenamefont{Pradhan and
  Chakrabarti}(2003{\natexlab{a}})}]{chakrabarti_fatigue}
\bibinfo{author}{\bibfnamefont{S.}~\bibnamefont{Pradhan}} \bibnamefont{and}
  \bibinfo{author}{\bibfnamefont{B.~K.} \bibnamefont{Chakrabarti}},
  \bibinfo{journal}{Phys. Rev. E} \textbf{\bibinfo{volume}{67}},
  \bibinfo{pages}{046124} (\bibinfo{year}{2003}{\natexlab{a}}).

\bibitem[{\citenamefont{Bhattacharyya et~al.}(2003)\citenamefont{Bhattacharyya,
  Pradhan, and Chakrabarti}}]{chakrabarti_phasetrans}
\bibinfo{author}{\bibfnamefont{P.}~\bibnamefont{Bhattacharyya}},
  \bibinfo{author}{\bibfnamefont{S.}~\bibnamefont{Pradhan}}, \bibnamefont{and}
  \bibinfo{author}{\bibfnamefont{B.~K.} \bibnamefont{Chakrabarti}},
  \bibinfo{journal}{Phys. Rev. E} \textbf{\bibinfo{volume}{67}},
  \bibinfo{pages}{046122} (\bibinfo{year}{2003}).

\bibitem[{\citenamefont{Batrouni et~al.}(2002)\citenamefont{Batrouni, Hansen,
  and Schmittbuhl}}]{batrouni_intfail_pre_2002}
\bibinfo{author}{\bibfnamefont{G.~G.} \bibnamefont{Batrouni}},
  \bibinfo{author}{\bibfnamefont{A.}~\bibnamefont{Hansen}}, \bibnamefont{and}
  \bibinfo{author}{\bibfnamefont{J.}~\bibnamefont{Schmittbuhl}},
  \bibinfo{journal}{Phys. Rev. E} \textbf{\bibinfo{volume}{65}},
  \bibinfo{pages}{036126} (\bibinfo{year}{2002}).

\bibitem[{\citenamefont{Delaplace et~al.}(1999)\citenamefont{Delaplace, Roux,
  and Pijaudier-Callot}}]{delaplace_ijss_1999}
\bibinfo{author}{\bibfnamefont{A.}~\bibnamefont{Delaplace}},
  \bibinfo{author}{\bibfnamefont{S.}~\bibnamefont{Roux}}, \bibnamefont{and}
  \bibinfo{author}{\bibfnamefont{G.}~\bibnamefont{Pijaudier-Callot}},
  \bibinfo{journal}{Int. J. Solids Struct.} \textbf{\bibinfo{volume}{36}},
  \bibinfo{pages}{1403} (\bibinfo{year}{1999}).

\bibitem[{\citenamefont{Roux et~al.}(1999)\citenamefont{Roux, Delaplace, and
  Pijaudier-Cabot}}]{roux_damint_physicaa_1999}
\bibinfo{author}{\bibfnamefont{S.}~\bibnamefont{Roux}},
  \bibinfo{author}{\bibfnamefont{A.}~\bibnamefont{Delaplace}},
  \bibnamefont{and}
  \bibinfo{author}{\bibfnamefont{G.}~\bibnamefont{Pijaudier-Cabot}},
  \bibinfo{journal}{Physica A} \textbf{\bibinfo{volume}{270}},
  \bibinfo{pages}{35} (\bibinfo{year}{1999}).

\bibitem[{\citenamefont{Zapperi et~al.}(2000)\citenamefont{Zapperi, Herrmann,
  and Roux}}]{zapperi_crackfuse_eujb_2000}
\bibinfo{author}{\bibfnamefont{S.}~\bibnamefont{Zapperi}},
  \bibinfo{author}{\bibfnamefont{H.~J.} \bibnamefont{Herrmann}},
  \bibnamefont{and} \bibinfo{author}{\bibfnamefont{S.}~\bibnamefont{Roux}},
  \bibinfo{journal}{Eur. Phys. J. B} \textbf{\bibinfo{volume}{17}},
  \bibinfo{pages}{131} (\bibinfo{year}{2000}).

\bibitem[{\citenamefont{Knudsen and
  Massih}(2004)}]{knudsen_breaksurf_condmat_2004}
\bibinfo{author}{\bibfnamefont{J.}~\bibnamefont{Knudsen}} \bibnamefont{and}
  \bibinfo{author}{\bibfnamefont{A.~R.} \bibnamefont{Massih}}
  (\bibinfo{year}{2004}), \bibinfo{note}{cond-mat/0410280}.

\bibitem[{\citenamefont{Herrmann et~al.}(1989)\citenamefont{Herrmann, Hansen,
  and Roux}}]{hjh_prb_fractdis_1989}
\bibinfo{author}{\bibfnamefont{H.~J.} \bibnamefont{Herrmann}},
  \bibinfo{author}{\bibfnamefont{A.}~\bibnamefont{Hansen}}, \bibnamefont{and}
  \bibinfo{author}{\bibfnamefont{S.}~\bibnamefont{Roux}},
  \bibinfo{journal}{Phys. {R}ev. {B}} \textbf{\bibinfo{volume}{39}},
  \bibinfo{pages}{637} (\bibinfo{year}{1989}).

\bibitem[{\citenamefont{Landau and Lifschitz}(1986)}]{ll_elast_en}
\bibinfo{author}{\bibfnamefont{L.~D.} \bibnamefont{Landau}} \bibnamefont{and}
  \bibinfo{author}{\bibfnamefont{E.~M.} \bibnamefont{Lifschitz}},
  \emph{\bibinfo{title}{Theory of elasticity}}
  (\bibinfo{publisher}{Butterworth-{H}einemann}, \bibinfo{year}{1986}),
  \bibinfo{edition}{3rd} ed.

\bibitem[{\citenamefont{Beitz and Grote}(2001)}]{dubbel_20}
\bibinfo{editor}{\bibfnamefont{W.}~\bibnamefont{Beitz}} \bibnamefont{and}
  \bibinfo{editor}{\bibfnamefont{K.-H.} \bibnamefont{Grote}}, eds.,
  \emph{\bibinfo{title}{Dubbel Taschenbuch f{\"{u}}r den {M}aschinenbau}}
  (\bibinfo{publisher}{Springer-{V}erlag}, \bibinfo{address}{New {Y}ork and
  others}, \bibinfo{year}{2001}), \bibinfo{edition}{20th} ed.

\bibitem[{\citenamefont{Sornette}(1989{\natexlab{b}})}]{sornette_jp_1989}
\bibinfo{author}{\bibfnamefont{D.}~\bibnamefont{Sornette}},
  \bibinfo{journal}{J. Phys. (France)} \textbf{\bibinfo{volume}{50}},
  \bibinfo{pages}{745} (\bibinfo{year}{1989}{\natexlab{b}}).

\bibitem[{\citenamefont{Pradhan and
  Chakrabarti}(2003{\natexlab{b}})}]{pradhan_ijmpb_2003}
\bibinfo{author}{\bibfnamefont{S.}~\bibnamefont{Pradhan}} \bibnamefont{and}
  \bibinfo{author}{\bibfnamefont{B.~K.} \bibnamefont{Chakrabarti}},
  \bibinfo{journal}{Int. {J}. {M}od. {P}hys. B} \textbf{\bibinfo{volume}{17}},
  \bibinfo{pages}{5565} (\bibinfo{year}{2003}{\natexlab{b}}).

\bibitem[{\citenamefont{Hemmer and Hansen}(1992)}]{hemmer_distburst_jam_1992}
\bibinfo{author}{\bibfnamefont{P.~C.} \bibnamefont{Hemmer}} \bibnamefont{and}
  \bibinfo{author}{\bibfnamefont{A.}~\bibnamefont{Hansen}},
  \bibinfo{journal}{J. Appl. Mech.} \textbf{\bibinfo{volume}{59}},
  \bibinfo{pages}{909} (\bibinfo{year}{1992}).

\bibitem[{\citenamefont{Hansen and Hemmer}(1994)}]{hansen_distburst_local_1994}
\bibinfo{author}{\bibfnamefont{A.}~\bibnamefont{Hansen}} \bibnamefont{and}
  \bibinfo{author}{\bibfnamefont{P.~C.} \bibnamefont{Hemmer}},
  \bibinfo{journal}{Phys. Lett. A} \textbf{\bibinfo{volume}{184}},
  \bibinfo{pages}{394} (\bibinfo{year}{1994}).

\bibitem[{\citenamefont{Andersen et~al.}(1997)\citenamefont{Andersen, Sornette,
  and Leung}}]{andersen_tricrup_prl_1997}
\bibinfo{author}{\bibfnamefont{J.}~\bibnamefont{Andersen}},
  \bibinfo{author}{\bibfnamefont{D.}~\bibnamefont{Sornette}}, \bibnamefont{and}
  \bibinfo{author}{\bibfnamefont{K.}~\bibnamefont{Leung}},
  \bibinfo{journal}{Phys. Rev. Lett.} \textbf{\bibinfo{volume}{78}},
  \bibinfo{pages}{2140} (\bibinfo{year}{1997}).

\bibitem[{\citenamefont{{{D'} Addetta} et~al.}(2001)\citenamefont{{{D'}
  Addetta}, Kun, Ramm, and Herrmann}}]{HJH_cont_and_discont_2000}
\bibinfo{author}{\bibfnamefont{G.~A.} \bibnamefont{{{D'} Addetta}}},
  \bibinfo{author}{\bibfnamefont{F.}~\bibnamefont{Kun}},
  \bibinfo{author}{\bibfnamefont{E.}~\bibnamefont{Ramm}}, \bibnamefont{and}
  \bibinfo{author}{\bibfnamefont{H.~J.} \bibnamefont{Herrmann}}, in
  \emph{\bibinfo{booktitle}{{Continuous} and {D}iscontinuous {M}odelling of
  {C}ohesive-{F}rictional {M}aterials}}, edited by
  \bibinfo{editor}{\bibfnamefont{P.}~\bibnamefont{Vermeer}}
  \bibnamefont{et~al.} (\bibinfo{publisher}{Springer-{V}erlag {B}erlin
  {H}eidelberg {N}ew {Y}ork}, \bibinfo{year}{2001}), Lecture {N}otes in
  {P}hysics.

\end{thebibliography}
\end{document}